\documentclass[conference]{IEEEtran}

\usepackage{cite}
\usepackage{amsmath,amssymb,amsfonts,amsthm}
\usepackage{algorithmic}
\usepackage{graphicx}
\usepackage{textcomp}
\usepackage{xcolor}
\def\BibTeX{{\rm B\kern-.05em{\sc i\kern-.025em b}\kern-.08em
    T\kern-.1667em\lower.7ex\hbox{E}\kern-.125emX}}
    
\DeclareMathOperator*{\argmin}{argmin}
\DeclareMathOperator*{\argmax}{argmax}

\theoremstyle{definition}
\newtheorem{definition}{Definition}
 
\theoremstyle{remark}
\newtheorem*{remark}{Remark}

\newtheorem{hyp}{Hypothesis}
\newtheorem{alg}{Algorithm}

\usepackage{url}
\usepackage{tikz}
\usetikzlibrary{bayesnet}
\usepackage{mathrsfs}

\usepackage[english]{babel}

\begin{document}

\title{Characterizing Entities in the  Bitcoin Blockchain}

\author{\IEEEauthorblockN{Marc Jourdan}
\IEEEauthorblockA{\textit{IBM Research, Singapore}\\
10 Marina Boulevard \\
18983 Singapore \\
mjourdan@sg.ibm.com}
\and
\IEEEauthorblockN{Sebastien Blandin}
\IEEEauthorblockA{\textit{IBM Research, Singapore}\\
10 Marina Boulevard \\
18983 Singapore \\
sblandin@sg.ibm.com}
\and
\IEEEauthorblockN{Laura Wynter}
\IEEEauthorblockA{\textit{IBM Research, Singapore}\\
10 Marina Boulevard \\
18983 Singapore \\
lwynter@sg.ibm.com}
\and
\IEEEauthorblockN{Pralhad Deshpande}
\IEEEauthorblockA{\textit{IBM Research, Singapore}\\
10 Marina Boulevard \\
18983 Singapore \\
pralhad@sg.ibm.com}
}

\maketitle

\begin{abstract}
Bitcoin has created a new exchange paradigm within which financial transactions can be trusted without an intermediary. This premise of a free decentralized transactional network however requires, in its current implementation, unrestricted access to the ledger for peer-based transaction verification. A number of studies have shown that, in this pseudonymous context, identities can be leaked based on transaction features or off-network information. In this work, we analyze the information revealed by the pattern of transactions in the neighborhood of a given entity transaction. By definition, these features which pertain to an extended network are not directly controllable by the entity, but might enable leakage of information about transacting entities. We define a number of new features relevant to entity characterization on the Bitcoin Blockchain and study their efficacy in practice.  We show that even a weak attacker with shallow data mining knowledge is able to leverage these features to characterize the entity properties.
\end{abstract}
\begin{section}{Introduction}
 Bitcoin~\cite{Nakamoto} stands out as the first global decentralized currency, and has seen spectacular growth recently, as illustrated by the exponential shape of the value of a transaction fee over the year $2017$, see Figure~\ref{fig:txFees}.  
\begin{figure}[!htb]
    \centering
    \includegraphics[scale=0.25]{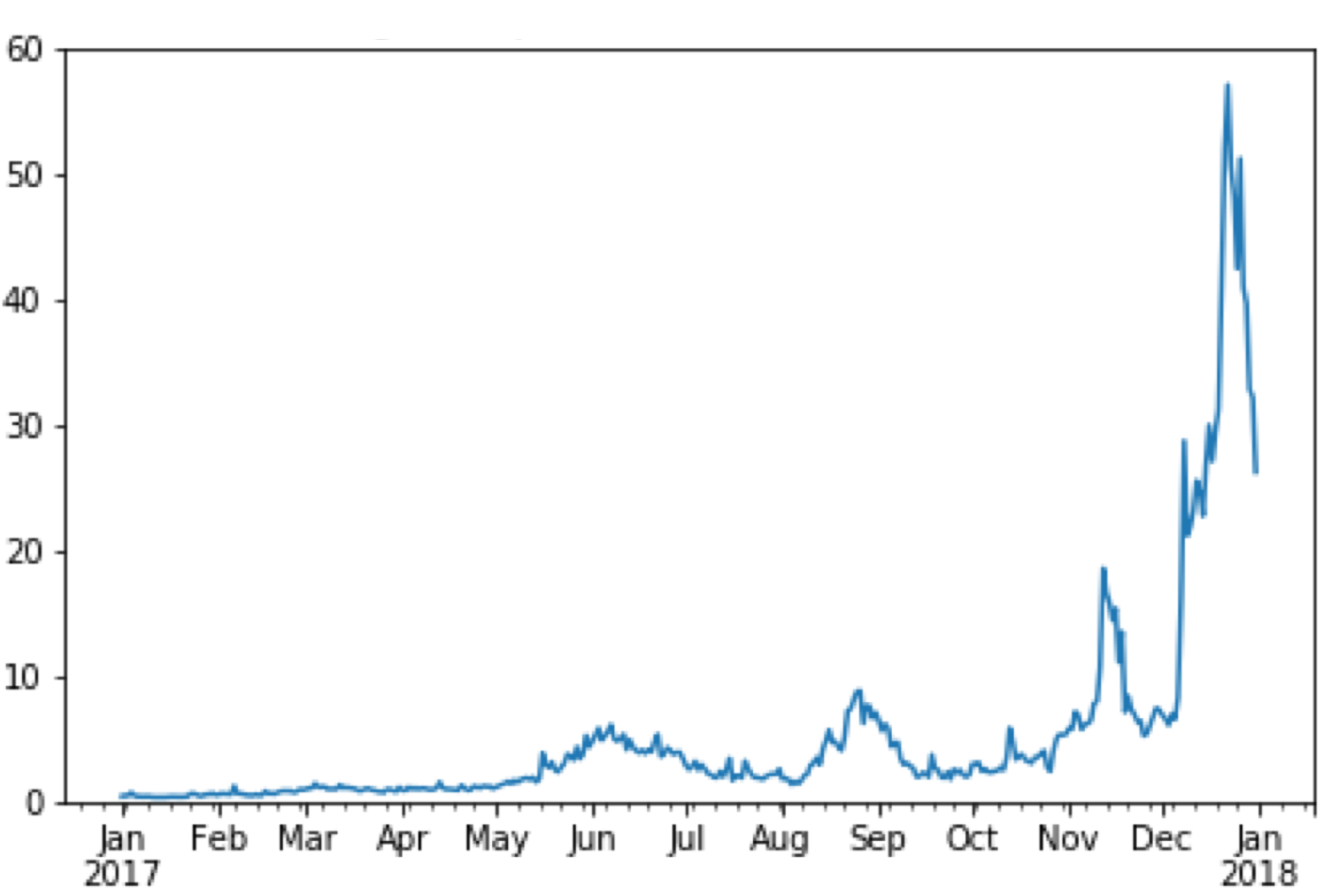}
    \caption{\textbf{Transaction fee in USD:} over the year $2017$.}
    \label{fig:txFees}
\end{figure}
The underlying Bitcoin data structure, the Blockchain, has been perceived as a catalyst to the emergence of broad decentralized applications, from crypto-currency exchanges, to decentralized autonomous organizations (DAO), or tokens, see~\cite{GraphPrimerBlockchain} for a comprehensive review. However, one of the most compelling applications remains the promise of a global decentralized currency, supporting  large portions of the global economy.
\begin{subsection}{Emergence of a decentralized global currency}
As presented in~\cite{FirstFourYear}, after only a few years, the bitcoin network has emerged to reflect a complex global payment system with new forms of players representing the traditional actors of the established financial infrastructure. Global transactions between Bitcoin exchanges, illustrated in Figure~\ref{fig:emergence_Tx_Net}, have drastically complexified in the last $5$ years.
\begin{figure}[!htb]
    \centering
    \includegraphics[scale=0.12]{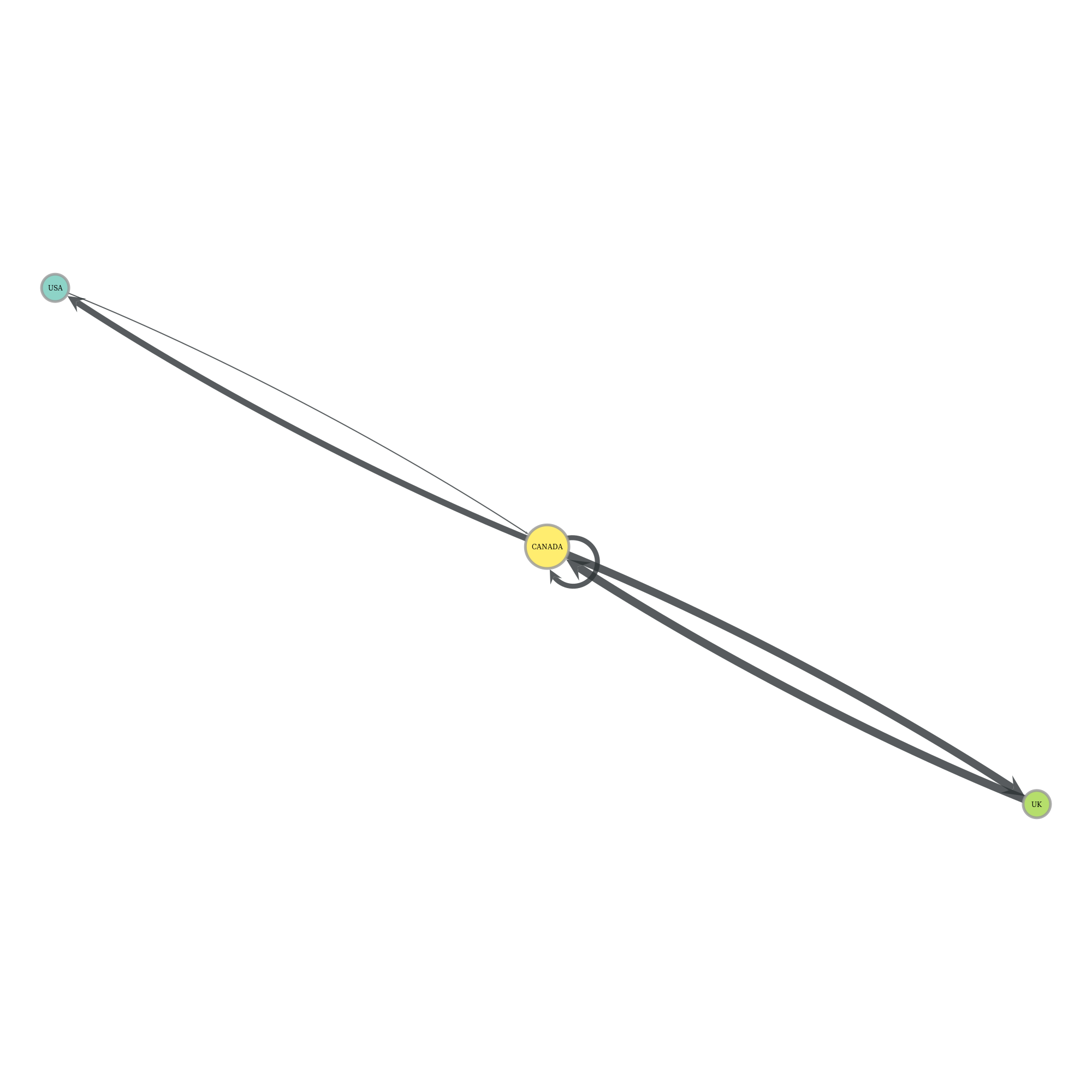}
    \includegraphics[scale=0.12]{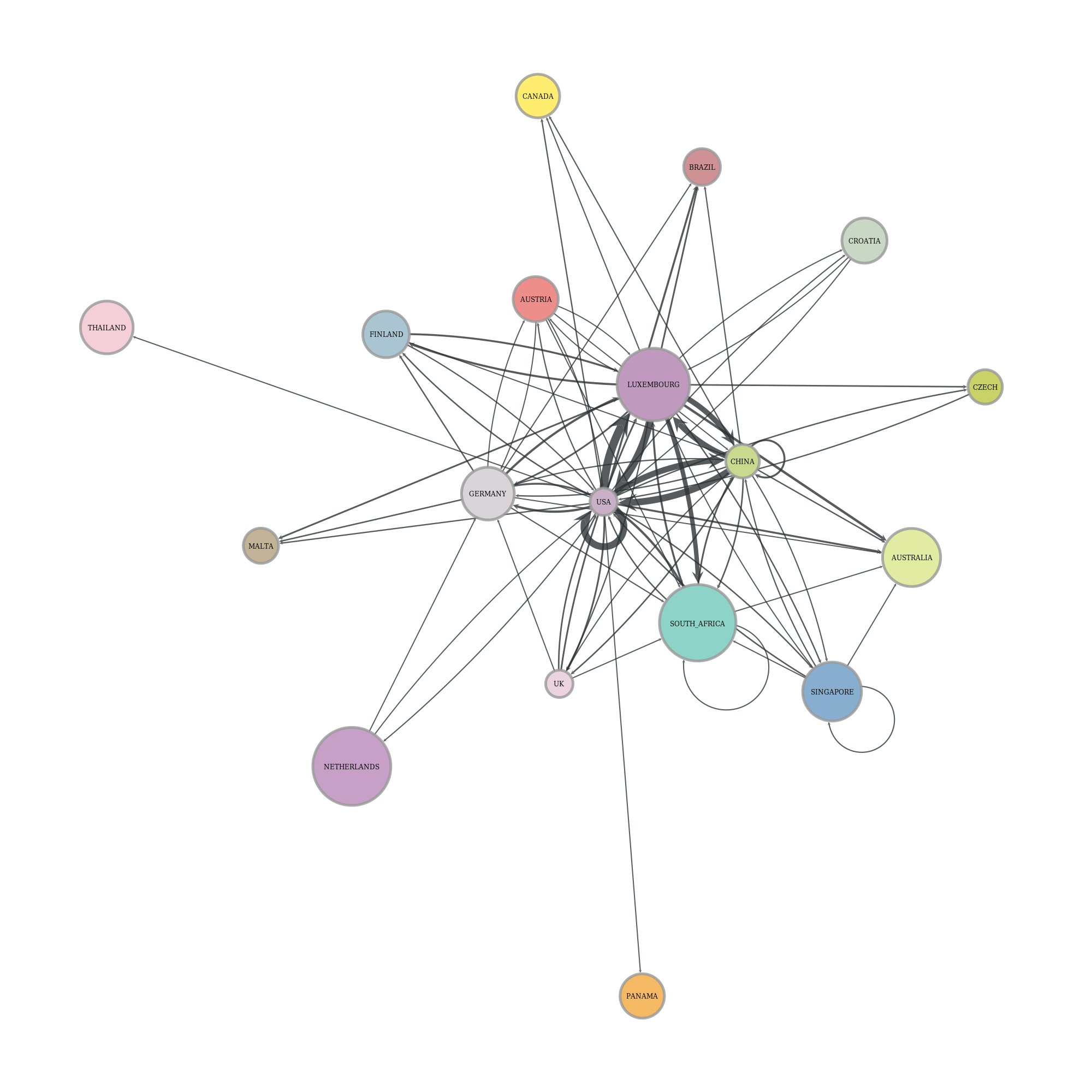}
    \caption{\textbf{Bitcoin exchanges transactions:} for the month of March $2013$ (top) and March $2018$ (bottom). The width of the edge is proportional to the total value of transactions during the month for the associated exchange pair (the size of the nodes is arbitrary).}
    \label{fig:emergence_Tx_Net}
\end{figure}
It is important for regulatory purposes as well as for the development of new crypto-currency paradigms to understand Bitcoin anonymity properties. Indeed, a healthy economy requires application of the \textit{rule of law} on financial transactions~\cite{MoneyLaundering}, which usually entails traceability and transparency on identities. On the other hand, in order to attract users, privacy considerations associated with crypto-currency have to be adequate and competitive with privacy properties of transactions using fiat money.

We say that a transaction medium is \textit{private} if no information on the transacting entity is revealed by the transaction graph. In the Bitcoin context, each individual transaction is publicly associated with a pseudonymous identity. Hence the strongest possibly guarantee is that the set of transactions associated with a given pseudonymous identity does not reveal information about the hidden transacting entity.

The main contribution of this work is the illustration that patterns of transactions involving the transacting entity but also patterns of neighboring transactions (on the transaction graph) are characteristic of the entity, in that these patterns can be turned into a fingerprint of transacting entity classes. We further give evidence that using these neighboring transaction patterns allows reaching state-of-the-art entity classification results.
\end{subsection}

\begin{subsection}{Related work}
Because the Bitcoin transaction graph is completely public it is clear that Bitcoin flows can be traced, although this is limited by merging transactions~\cite{QuantitativeAnalysisBTCtx}. Recent analysis of ransomware attacks and associated bitcoin transactions can be found in ~\cite{meiklejohn2013fistful,huang2018tracking,paquet2018ransomware}. 

Furthermore, joining Bitcoin transaction with off-network activities can contribute to linking identities of certain Bitcoin transactions~\cite{ReidHarrigan}, for instance the authors of~\cite{TorDeanonimization} link Bitcoin users with Tor hidden services.

A number of studies are focused on the problem of \textit{address clustering} consisting of identifying the address set associated with a transacting entity, and typically rely on heuristics motivated by properties of the Bitcoin Blockchain protocol, such as the requirement of a unique signature for transaction inputs, see for instance~\cite{AutomaticBtcClustering,7796940} and references therein for more details. The authors of~\cite{UnreasonableEffectiveness} highlight that these methods crucially depend on address re-use behavior.

Another attack motivated by the Bitcoin protocol involves the inference of peer-to-peer communication structure~\cite{NIPS2017_6735} and information leakage from the message dissemination pattern. A different type of attack is presented in~\cite{nick2015data}, where the authors show that statistical analysis of bloom filters could help to identify the set of addresses owned by Bitcoin wallets. It has also been shown that patterns of newly minted bitcoins could play a role in revealing information of certain Bitcoin users~\cite{mcginn2018toward}.

Thematically closer to our work, several studies have  considered the extent to which data mining methods can be applied to the entity characterization problem, with for instance the use of transaction-specific features in~\cite{multiClass}, able to achieve $70\%$ accuracy for classifying entities into several types. In ~\cite{ExchangeAddressClustering}, the authors introduce the notion of transaction motifs with application to the detection of bitcoin exchanges, and achieve greater than $80\%$ accuracy.
\end{subsection}
\begin{subsection}{Contributions of this work}
Motivated by the success of~\cite{ExchangeAddressClustering}, we consider the more general problem of classifying entities into multiple classes, based on extended transaction neighborhood properties. The study of transaction graph neighborhood structure is promising for at least two reasons.

First, the use of  graph network features has been shown to be  efficient for graph learning problems~\cite{verma2017hunt}. The promise of such methods is illustrated by the spread of graph databases and related applications~\cite{tanase2018system}. The authors of~\cite{narayanan2011link} for example are able to successfully re-identify nodes from a noisy graph structure provided as part of a Kaggle contest.

Second, if neighboring network transactions are confirmed to be informative of entity identities, they constitute a fundamental limit to Bitcoin privacy guarantees. Indeed, while an individual has control of his personal address usage and transaction patterns, he does not have  control of the behavior of the entities he is transacting with. While services such as CoinJoin serve to limit information leakage of Bitcoin flows via the merging of individual transactions~\cite{priceofanonymity}, there is at this stage no service providing full neighborhood obfuscation.

The main contribution of this work are the following:
\begin{itemize}
\item definition of novel features for entity classification from a graph neighborhood perspective,
\item analysis of the performance of various classification methods,
\item discussion of the implications of these entity characterization results in the context of Bitcoin anonymity.
\end{itemize} 
The structure of the paper is as follows. In Section~\ref{sec:graph} we define the graph model we propose for the Bitcoin Blockchain. In Section~\ref{sec:class}, we present our classification models as well as details of the graph neighborhood features developed. Finally we provide in Section~\ref{sec:numRes} numerical results that demonstrate the effectiveness of entity characterization using actual Bitcoin Blockchain data. Section~\ref{sec:conc} provides concluding remarks.
\end{subsection}
\end{section}
\begin{section}{Blockchain graph model}\label{sec:graph}
The Bitcoin Blockchain is a succession of blocks $\mathscr{B}$. Each block $b \in \mathscr{B}$ contains a set of transactions $T(b)=\{t_{i}\}_{i} \subset \mathscr{T} $. We first present a bipartite address-transaction graph model, and then explain how we derive a discrete-time entity-transaction graph model with features relevant to the entity characterization program.

\begin{subsection}{Address-transaction graph}

We model the Bitcoin Blockchain as a directed weighted bipartite graph $\mathscr{H} = (\mathscr{A}, \mathscr{T}, \mathscr{L})$, where $a \in \mathscr{A}$ represents an address, $t \in \mathscr{T}$ represents a transaction between addresses, and $l \in \mathscr{L}$ represents an edge between an address and a transaction. We partition the edge set into edges incoming to a transaction, and edges outgoing from a transaction, as follows: $l \in \mathscr{L} = \mathscr{I} \cup \mathscr{O}$, where $\mathscr{I}$ stands for input edges, i.e. incoming edges of a vertex $t$, and $\mathscr{O}$ for output edges, i.e. outgoing edges of a vertex $t$. 
Multiple edges can exist between a pair $(a,t) \in \mathscr{A} \times \mathscr{T}$. 

For each transaction $t \in \mathscr{T}$, we define $I(t) \subset \mathscr{I}$ and $O(t) \subset \mathscr{O}$ as the corresponding edges between the transaction and the addresses associated. For each $i \in \mathscr{I}$ (resp. $o \in \mathscr{O}$), we  uniquely define the associated transaction  $t(i)$ (resp. $t(o)$) and  address  $a(i)$ (resp. $a(o)$); these are well defined because the graph is bipartite. 

We ignore the address and transaction fields that are specific to the protocol or not relevant to our study, see for instance~\cite{HyperledgerFabric} for related graph models for permissioned blockchains, or~\cite{Cachin2017g} for a more fundamental analysis of a transaction graph model and inherited protocol properties. In this work, we consider the following vertex and edge properties:
\begin{itemize}
\item \textit{edge}: inputs, $\mathscr{I}$, (resp.  outputs, $\mathscr{O}$) are represented by their amount in BTC, $v(i)$ for $i \in \mathscr{I}$ ($v(o)$ for $o \in \mathscr{O}$), sent (resp. received) by an address, $a(i)$ (resp. $a(o)$), for a given transaction, $t(i)$ (resp. $t(o)$),
\item \textit{transaction vertex}: a transaction contained in a block $b(t)$ has a fee, $f(t)$, and a time $\tau(t)$, corresponding to the time when the block it belongs to is validated, $\tau(t)=\tau(b(t))$. 
\item \textit{address vertex}: an address has a creation date and a balance.
\end{itemize}
In the next section we explain how we derive the entity-transaction graph from the address-transaction graph.

\end{subsection}

\begin{subsection}{Entity-transaction graph}

In the Bitcoin Blockchain a  user  may employ several addresses. We therefore introduce the concept of an ``entity'' where  entity, $e$, is fully characterized by a set of addresses $A(e) = \{a^{(e)}_{i}\}_{i}$, which can be interpreted as a logical  user. We discuss subsequently the various methods for defining these sets, such as the common spending heuristic.

The entity-transaction graph is a directed weighted bipartite graph $ \mathscr{G} = (\mathscr{E}, \mathscr{T}_{e}, \mathscr{L}_{e})$ of entities and associated transactions. Let $e \in \mathscr{E}$ denote an entity and $t \in \mathscr{T}_{e}$,  a transaction between entities. Entities and transactions are connected by edges $l \in \mathscr{L}_{e} = \mathscr{I}_{e} \cup \mathscr{O}_{e}$, where $\mathscr{I}_{e}$ stands for input edges,  and $\mathscr{O}_{e}$ represents output edges.

As with addresses, for each transaction $t \in \mathscr{T}_{e}$ we define $I_{e}(t) \subset \mathscr{I}_{e}$ and $O_{e}(t) \subset \mathscr{O}_{e}$ as the corresponding edges between the transactions and  entities. For each $i \in \mathscr{I}_{e}$ (resp. $o \in \mathscr{O}_{e}$), as for the address-transaction graph, we uniquely define the associated transaction  $t(i)$ (resp. $t(o)$) and  entity  $e(i)$ (resp. $e(o)$). Inputs, $\mathscr{I}_{e}$, (resp. outputs, $\mathscr{O}_{e}$) represent the amount in BTC, $v(i)$ for $i \in \mathscr{I}_{e}$ ($v(o)$ for $o \in \mathscr{O}_{e}$), sent (resp. received) by an entity, $e(i)$ (resp. $e(o)$), for a given transaction. We define  the set of addresses, $A(I_{e}(t))$ (resp. $A(O_{e}(t))$), associated with a set of inputs (resp. outputs). The mechanism to build the entity-transaction graph from the address-transaction graph is illustrated in Figure~\ref{fig:EntityGraph}.

\begin{figure}[htb!]
\centering
\scalebox{0.7}{\begin{tikzpicture}[node distance=1.3cm,bend angle=20,auto]

  \tikzstyle{place}=[circle,thick,draw=blue!75,fill=blue!20,minimum size=10mm]
  \tikzstyle{redplace}=[circle,thick,draw=red!75,fill=red!20,minimum size=10mm]
  \tikzstyle{transition}=[rectangle,thick,draw=black!75,fill=black!20,minimum size=6mm]
        \begin{scope}
    \node [place] (n1) at (0,6.5) {$a^{(e_{1})}_{2}$};
    \node [place] (n2) [above of=n1]  {$a^{(e_{1})}_{1}$};
    \node [place] (n3) [below of=n1]  {$a^{(e_{1})}_{3}$};
    
    \node [place] (n4) at (4,6.5) {$a^{(e_{2})}_{3}$};
    \node [place] (n5) [above of=n4]  {$a^{(e_{2})}_{2}$};
    \node [place] (n6) [above of=n5]  {$a^{(e_{2})}_{1}$};
    \node [place] (n7) [below of=n4]  {$a^{(e_{3})}_{1}$};
    \node [place] (n8) [below of=n7]  {$a^{(e_{1})}_{4}$};
    
    \node [transition] (e1) at (2,6.5) {$t$};
      
    \draw[->]
        (n1) edge (e1) (e1) edge (n4) (n2) edge (e1) (e1) edge (n5) (n3) edge (e1) (e1) edge (n5) (e1) edge (n6) (e1) edge (n7) (e1) edge (n8);
    \draw[->] (e1) edge[bend left] (n6);
    
  \end{scope}
  
  \begin{scope}[xshift=7cm]
    \node [redplace] (n1') at (0,6.5) {$e_{1}$};
    
    \node [redplace] (n4') at (4,6.5) {$e_{3}$};
    \node [redplace] (n5') [above of=n4']  {$e_{2}$};
        \node [redplace] (n6') [below of=n4']  {$e_{1}$};
    
    \node [transition] (e1') at (2,6.5) {$t$};
      
    \draw[->]  (n1') edge (e1') (e1') edge (n4') (e1') edge (n5') (e1') edge (n5') (e1') edge (n6');
        
  \end{scope}
 
\end{tikzpicture}}
\caption{\textbf{Entity graph}: (right) is obtained by aggregation of the address graph (left).} 
\label{fig:EntityGraph}
\end{figure}
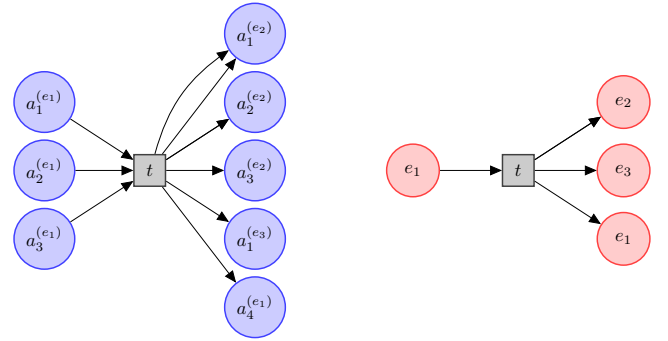
The vertices and edges of the entity-transaction graph inherit the properties of the address-transaction graph by standard summation of continuous properties and aggregation of graph topology. Additionally we consider that vertices and edges of the entity-transaction graph inherit some statistics of the entity-transaction graph that are lost by summation (e.g. an entity graph edge inherits the count of address graph edges which it subsumes).

We further define a set of categories $c \in \mathscr{C}$, corresponding to entity class labels. We shall consider in the remainder of this work the entity classification problem, namely identification of the class label $c \in \mathscr{C}$ associated with an entity $e \in \mathscr{E}$, based on the address-transaction graph.
\end{subsection}


\begin{subsection}{Discrete time graph} \label{Time based aggregated graphs}
We apply temporal aggregation to the directed weighted bipartite graph of entities $\mathscr{G} = (\mathscr{E},\mathscr{T}_{e},\mathscr{L}_{e})$ to obtain a discrete-time data structure with commensurate properties.
\begin{definition}[Discrete-time operator]
A discrete-time aggregation operator is an operator :
\[
\Delta^{(\tau_{1},\tau_{2})} \colon \mathscr{G} \mapsto \mathscr{G}^{(\tau_{1},\tau_{2})}
\]
where time-aggregated entities and transactions $\mathscr{E}^{(\tau_{1},\tau_{2})},\mathscr{T}_{e}^{(\tau_{1},\tau_{2})}$ are derived from $\mathscr{E},\mathscr{T}_{e},\mathscr{L}_{e}$ and required to satisfy the following constraints:
\begin{itemize}
    \item Entity activity: only entities transacting during the time period are considered. $\mathscr{E}^{(\tau_{1},\tau_{2})} = \{e \in \mathscr{E}, \exists t \in \mathscr{T}_{e}, \tau(t) \in [\tau_{1},\tau_{2}] \land ((e,t) \in I_{e}(t) \lor (t,e) \in O_{e}(t))\}$
    \item Transaction: the edges, $(e_{i},e_{j}) \in \mathscr{T}_{e}^{(\tau_{1},\tau_{2})}$, represent the aggregation of  transactions between two entities within the time window, $\{ t \in \mathscr{T}_{e}, \tau(t) \in [\tau_{1},\tau_{2}] \land e(I_{e}(t))=e_{i} \land \exists o \in O_{e}(t), e(o)=e_{j} \}$. 
    
\end{itemize}
In the following, we work with the discrete-time graph $\mathscr{G}^{(\tau_{1},\tau_{2})} = (\mathscr{E}^{(\tau_{1},\tau_{2})},\mathscr{T}_{e}^{(\tau_{1},\tau_{2})})$ obtained by applying the discrete-time operator, with typical discrete time intervals of a day, a week, or a month.
\end{definition}
\end{subsection}

\begin{subsection}{Motifs} \label{Motifs}

The notion of \textit{motif} in the Bitcoin Blockchain was introduced in~\cite{ExchangeAddressClustering} as a  useful concept in entity classification studies, specifically in the context of Exchange detection. The authors define the so-called 2-motif. In this work, we  consider more generally the case of N-motif and present a few relevant special cases.

\begin{definition}[1-motif]
A 1-motif is a path of length 2 on the entity-transaction graph, $(e_{1},t,e_{2}) \in \mathscr{E} \times \mathscr{T}_{e} \times \mathscr{E}$, in $\mathscr{G}$.
\begin{equation*}
motif_{1}(t) = \{(e_{i},t,e_{j})\}_{i,j}. 
\end{equation*}
If $e_{1} = e_{2}$ we call it a Loop 1-motif, otherwise it is called  Distinct 1-motif.
\end{definition}
The 1-motif, is a direct transaction between entities. More generally, a direct N-motif is a path of length $2N$ in the directed weighted bipartite graph starting and ending with an entity. 
\begin{definition}[Direct N-motif]
A N-motif is a path of length $2N$, $(e_{1},t_{1},\dots,t_{N},e_{N+1}) \in \mathscr{E} \times \mathscr{T}_{e} \times \dots \times \mathscr{T}_{e} \times \mathscr{E}$, in $\mathscr{G}$ starting and ending with an entity.
A motif$_{N}(t_{1},\dots,t_{N})=\{(e_{i_{1}},t_{1},\dots,t_{N},e_{i_{N+1}})\}$ is required to satisfy the following constraint:
\begin{itemize}
    \item Direct: at least one output from each transaction is an input to the next transaction. $\forall k \in \{1, \dots, N-1\} \exists (o,i) \in O(t_{k}) \times I(t_{k+1}), a(o)=a(i)$. 
\end{itemize}
If $e_{1} = e_{N+1}$ we call it a Direct Loop, otherwise Direct Distinct. From this condition we deduce immediately that the transactions are ordered in time: $\tau(t_{1}) < \dots < \tau(t_{N})$. Considering only Direct N-motif avoids redundancy of exploration and focuses on fast flow of value. In this work, we do not consider non-Direct paths that would correspond to more complex transfer-and-hold patterns.

\end{definition}
We illustrate the case of a 3-motif in Figure~\ref{fig:3motif}.
\begin{figure}[htb]
    \centering
    \scalebox{0.65}{\begin{tikzpicture}[node distance=2cm,bend angle=30,auto]

    \tikzstyle{place}=[circle,thick,draw=blue!75,fill=blue!20,minimum size=10mm]
    \tikzstyle{transition}=[rectangle,thick,draw=black!75,fill=black!20,minimum size=4mm]
        \begin{scope}    
    \node [place] (n1) at (0,0) {$e_{1}$};
    \node [place] (n2) at (4,0) {$e_{2}$};
    \node [place] (n3) at (8,0) {$e_{3}$};
    \node [place] (n4) at (12,0) {$e_{4}$};
    \node [transition] (e1) at (2,0) {$t^{(3)}_{1}$};
    \node [transition] (e2) at (6,0) {$t^{(3)}_{2}$};
    \node [transition] (e3) at (10,0) {$t^{(3)}_{3}$};
    
    \draw[->] (n1) edge (e1) (e1) edge (n2) (n2) edge (e2) (e2) edge (n3) (n3) edge (e3) (e3) edge (n4);
    
        \end{scope}
 
    \end{tikzpicture}}
    \caption{\textbf{3-motif}: consisting of a path of length $6$ on the bipartite entity-transaction graph.}
    \label{fig:3motif}
\end{figure}
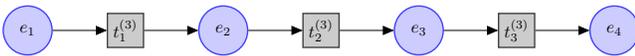

Statistics of 1,2, and 3-motifs over the dataset considered are presented in Table~\ref{table:motif_statistics}, with the following entity categories: Exchange, Gambling, Mining, Service, Darknet, as per labelling from Wallet Explorer. An indication of the power of transaction graph neighborhood structure is that certain motifs dominate within certain  entity  categories, e.g. direct transactions with distinct entities are more characteristic of Exchanges than of other entity categories. 
\begin{table*}[!htb]
\centering
\caption{\textbf{Motif distribution}:  1-motif Distinct is indicative of Exchanges ($64.6\%$ of such motifs correspond to Exchanges). On the other hand 3-motif Direct Loop is indicative of Services ($63.0\%$ of such motifs correspond to Services).}
\begin{tabular}{||c c c c c c c c||}
 \hline
 Type & Sub-type & Quantity & Exchange & Gambling & Mining & Service & Darknet\\ [0.5ex] 
 \hline\hline
 1-motif &  Loop  &  16.085.493 & 26.3\%  &  25.7\%  &   4.9\%  &  39.6\%  &  3.5\%\\
 1-motif &  Distinct  &   5.390.310 & 64.6\%  &  1.1\%  &  2.7\%  &  30.3\%  &  1.2\%\\
 2-motif & Direct Loop &  10.196.844  & 21.1\% &  28.1\%  &  0.1\%  &  48.5\%  &  2.1\%\\
 2-motif & Direct Distinct &  20.469.285  & 46.9\%  &  6.3\%  &  3.7\%  &  38.7\%  &  4.4\% \\
 3-motif & Direct Loop  & 30.914.975  & 24.6\%  &  11.5\%  &  0.1\%  &  63.0\%  &  0.9\% \\
 3-motif & Direct Distinct  & 85.822.858   & 54.0\%  &  4.4\%  &  3.6\%  &  34.1\%  &  3.9\% \\[1ex] 
 \hline
\end{tabular}
\label{table:motif_statistics}
\end{table*}

      
    
 
Motif attributes such as BTC volume and number of addresses are  similarly inherited from the address-transaction network, by summation and aggregation.


\end{subsection}

\end{section}


\begin{section}{Entity classification}\label{sec:class}

In this section we present the  methods used for inferring the  category associated with each entity, using our graph neighborhood features on the Bitcoin transaction graph. We first recall the heuristics used for associating distinct addresses to a single entity.

\begin{subsection}{Common spending heuristic for address clustering}\label{sec:commSpend}

The common spending heuristic consists of clustering addresses that are inputs to the same transaction with the same entity. Formally,
\begin{hyp}[Common Spending] \label{hyp:CS}
\[
\forall t \in \mathscr{T}, \quad \exists ! e \in \mathscr{E}, \quad \forall i \in I(t), a(i) \in A(e)
\]
\end{hyp}
The common spending heuristic is equivalent to the assumption that having access to multiple private keys at a given point in time is a defining property of an entity. Indeed, in order to submit a transaction $T$ to the Blockchain protocol,  the transaction must be signed,  implying using the private key of each address in the input set of the transaction, see Figure~\ref{fig:CSheuristic}.
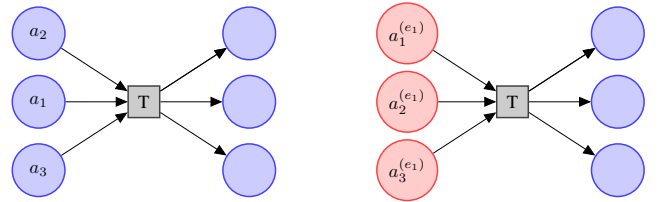
\begin{figure}[htb!]
    \centering
    \scalebox{0.7}{
\begin{tikzpicture}[node distance=1.3cm,bend angle=45,auto]

          \tikzstyle{place}=[circle,thick,draw=blue!75,fill=blue!20,minimum size=10mm]
          \tikzstyle{redplace}=[circle,thick,draw=red!75,fill=red!20,minimum size=10mm]
          \tikzstyle{transition}=[rectangle,thick,draw=black!75,fill=black!20,minimum size=6mm]
            \begin{scope}
                \node [place] (n1) at (0,6.5) {$a_{1}$};
                \node [place] (n2) [above of=n1]  {$a_{2}$};
                \node [place] (n3) [below of=n1]  {$a_{3}$};
                
                \node [place] (n4) at (4,6.5) {};
                \node [place] (n5) [above of=n4]  {};
                \node [place] (n7) [below of=n4]  {};
                
                \node [transition] (e1) at (2,6.5) {T};
                  
                \draw[->]
                    (n1) edge (e1) (e1) edge (n4) (n2) edge (e1) (e1) edge (n5) (n3) edge (e1) (e1) edge (n5)  (e1) edge (n7);
                
            \end{scope}
              
            \begin{scope}[xshift=7cm]
                \node [redplace] (n1') at (0,6.5) {$a^{(e_{1})}_{2}$};
                \node [redplace] (n2') [above of=n1']  {$a^{(e_{1})}_{1}$};
                \node [redplace] (n3') [below of=n1']  {$a^{(e_{1})}_{3}$};
                
                \node [place] (n4') at (4,6.5) {};
                \node [place] (n5') [above of=n4']  {};
                \node [place] (n7') [below of=n4']  {};
                
                \node [transition] (e1') at (2,6.5) {T};
                  
                \draw[->]  (n1') edge (e1') (n2') edge (e1') (n3') edge (e1') (e1') edge (n4') (e1') edge (n5') (e1') edge (n5') (e1') edge (n7');
            
            \end{scope}
 
        \end{tikzpicture}}
    \caption{\textbf{Common Spending Heuristic}: all addresses input to a transaction are associated with the same entity.}
    \label{fig:CSheuristic}
\end{figure}

The notion of transitive closure allows extending the set of addresses associated with a given entity.
\begin{hyp}[Transitive Closure] \label{hyp:TC}
\begin{equation*}
\begin{aligned}
\exists (t_{1},t_{2}) \in \mathscr{T}^{2}, s.t. \{a_{1},a_{2}\}=A(I(t_{1})), \{a_{1},a_{3}\}=A(I(t_{2})) \\
 \implies \exists ! e, \{a_{1},a_{2},a_{3}\} \subset A(e)
\end{aligned}
\end{equation*}
\end{hyp}

Under Hypotheses~\ref{hyp:CS},~\ref{hyp:TC}, each transaction only has one input entity: $\forall t \in \mathscr{T}_{e}, |I_{e}(t)|=1$. In the following section we present the types of features used for entity classification.
\end{subsection}

\begin{subsection}{Features}
We wish to demonstrate the effectiveness of  graph neighborhood features for entity characterization. As such we propose the following five feature classes:
\begin{itemize}
\item Address features involving only address properties,
\item Entity features that can be computed from the set of addresses associated with a given entity,
\item Temporal features related to the evolution of specific transaction properties,
\item Centrality features encoding the value of classical centrality measures~\cite{freeman1977set},
\item Motif features corresponding to transaction paths involving the entity of interest.
\end{itemize}
The remainder of this section  provides more details on each set of features. 

Address-specific features include attributes such as the total BTC received, the total BTC balance, the number of input/output transactions, the number of predecessor/successor addresses,  unique and otherwise, the number of predecessors addresses that are also successors, and the number of sibling addresses in output.

Analogous features are defined at the entity level as well as the number and proportion of coinbase transactions.

1-motif features  include the value of incoming/outgoing transactions in BTC and USD, the number of incoming/outgoing addresses, the number of incoming/outgoing transactions per day, their total value in BTC and USD, and their  total fee.

2-motif and 3-motif features are analogous but include also particularities of this graph structure such as, for 2-motif, the number of inputs (resp. outputs) in the first (resp. second) transaction of an incoming/outgoing/loop motif, the total value of the inputs (resp. outputs) of the first (resp. second) transaction of an incoming/outgoing/loop motif in BTC and USD, as well as the number of incoming/outgoing/loop motifs, the number of addresses involved as center of an incoming/outgoing/loop motif, the value transferred in the middle and the fees of the transactions in BTC and USD, and the number of predecessors/successors for an incoming/outgoing motif. 2-motif features are illustrated in Figure~\ref{fig:2motiffeature}.

\begin{figure}[htb]
    \centering
    \scalebox{0.65}{\begin{tikzpicture}[node distance=2cm,bend angle=30,auto]

    \tikzstyle{place}=[circle,thick,draw=blue!75,fill=blue!20,minimum size=10mm]
    \tikzstyle{transition}=[rectangle,thick,draw=black!75,fill=black!20,minimum size=4mm]
    \tikzstyle{label}=[rectangle,thick,draw=black!75,minimum size=4mm]
        \begin{scope}    
    \node [place] (n1) at (0,0) {$e_{1}$};
    \node [place] (n2) at (6,0) {$e_{2}$};
    \node [place] (n3) at (12,0) {$e_{3}$};
    \node [transition] (e1) at (3,0) {$t^{(3)}_{1}$};
    \node [transition] (e2) at (9,0) {$t^{(3)}_{2}$};
    
    \node[label] (l1) at (1.5,1) {nb\_inputs};
    \node[label] (l2) at (10.5,1) {nb\_outputs};
    \node[label] (l3) at (1.5,-1) {in\_val};
    \node[label] (l4) at (10.5,-1) {out\_val};
    \node[label] (l5) at (6,1) {nb\_address};
    
    \node[label] (l6) at (3,1) {Fee$_{1}$};
    \node[label] (l7) at (9,1) {Fee$_{2}$};

    \node[label] (l8) at (7.5,-1) {mid\_val};
    
    \draw[-] (l1) edge (l3) (l2) edge (l4)  (l6) edge (e1)  (l7) edge (e2) (l5) edge (n2);
      
    \draw[->] (n1) edge (e1) (e1) edge (n2) (n2) edge (e2) (e2) edge (n3) ;
    
        \end{scope}
 
    \end{tikzpicture}}
    \caption{\textbf{2-motif features} (rectangular white boxes) annotated over a 2-motif, including both edge features and vertex features over transaction paths.}
    \label{fig:2motiffeature}
\end{figure}
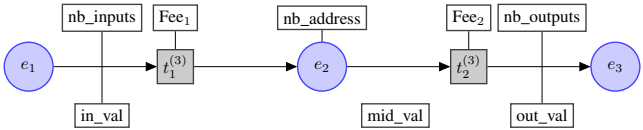

Centrality-based features include measures of  betweenness centrality, closeness centrality, degree centrality, in-degree centrality, out-degree centrality, PageRank, and load centrality. These features are computed on the discrete-time aggregation of the entity graph over a week, a month and a year.

Temporal features are those such as the number of  weeks, months, years of activity, the number of entity traded with per week, month, year, the number of receiving, sending days, the activity period duration, and the active day ratio.

We  make use of in total 10 address features, 8 entity features, 16 temporal features, 42 centrality features, 44 1-motif features, 81 2-motif features, and 114 3-motif features.
\begin{remark}
For each continuous feature such as volume of Bitcoin transactions, we calculate the mean and the standard deviation in order to allow the classifier to  discriminate up to second order statistics. 

Transaction values are considered both in terms of BTC and USD, in order to support the analysis over the multiple years over which the BTC value in USD changed.
\end{remark}
\end{subsection}

\begin{subsection}{Classification method}\label{sec:classMeth}

Given our interest in understanding the behavioral features characterizing certain categories of Bitcoin users, we propose to use a decision tree method, which provides feature relevance statistics via bootstrapping. A decision tree  $h_{m}(x)$ partitions the feature space into $J_{m}$ disjoint regions $R_{1m},\ldots ,R_{J_{m}m}$ and produces a constant value in each region. The output of $h_{m}(x)$ for input $x$ can be written as the sum:
\[
h_{m}(x)=\sum _{j=1}^{J_{m}}b_{jm}\mathbf {1} _{R_{jm}}(x)
\]
where $b_{jm}$ is the model response for input features from region $R_{jm}$. We consider an ensemble of trees, which we learn sequentially by minimizing the weighted sum 
\[F_{m}(x)=F_{m-1}(x)+\gamma _{m}h_{m}(x),\]
\[\gamma _{m}= \argmin_\gamma \sum_{i=1}^n L(y_i, F_{m-1}(x_i) + \gamma h_m(x_i)).
\]
according to a gradient boosting procedure, where $L(\cdot)$ is the loss function.

In order to identify the ensemble tree hyper-parameters most suited to our problem, we approximate the performance of the tree ensemble for given parameter values using Gaussian Processes. We learn a surrogate $\hat{f}(\cdot)$ of the loss function $f(\cdot)$ based on previous evaluations $\{(\theta_{1},f(\theta_{1})), \dots , (\theta_{k},f(\theta_{k})\}$, and identify the parameter minimizing this surrogate function. The calibration procedure can be summarized as follows:
\begin{alg}
For $t \in \{1, \dots, T\}$:
\begin{itemize}
    \item Given observations $\{(\theta_{i},f(\theta_{i})), \forall i \in \{1, \dots , t\} \}$, we build a surrogate $\hat{f}_{t}$ using Gaussian processes. Each value $f(\theta_{i})$ is the loss function value obtained after having trained a decision tree with the hyper-parameter $\theta_{i}$.
    \item Given the surrogate function $\hat{f}_{t}$, we identify the parameter $\theta_{t+1}$ providing a good compromise between minimizing the surrogate $\hat{f}_{t}$ and exploring the parameter space. The evaluation of the surrogate function for a parameter value consists in running the gradient boosting routine described above.
\end{itemize}
\end{alg}
We benchmark the result of our decision tree algorithm against a logistic regression algorithm, defined below:
\[
\forall x_{i} \in \mathscr{E},\quad \hat{f}_{entity}(x_{i}) = y_{i} = \argmax_{c \in \mathscr{C}} f_{c}(x_{i})
\]
where
\[
f_{c}(X) = \mathbb{P}(Y=c|X) = h((X^{\Phi})^{T}\beta_{c}),\quad h(t) = \frac{e^{t}}{1+e^{t}}.
\]

\end{subsection}
\end{section}

\begin{section}{Numerical Results}\label{sec:numRes}

In this section we present  our experimental results, as well as implications of these results for Bitcoin transaction anonymity.

\begin{subsection}{Blockchain dataset}

We consider the set of blocks of height inferior or equal to 514.971, corresponding to blocks created before March 24th 2018, 15:19:02, which contains about $500.000.000$ addresses. Address labels are obtained from WalletExplorer\footnote{\url{https://www.walletexplorer.com/}}.

We apply the common spending heuristic and transitive closure operation described in Section~\ref{sec:commSpend} to the labeled dataset obtained from WalletExplorer, and extend it slightly. We interact with the Bockchain via the BlockSci toolbox v.0.4.5 released on March 16th 2018~\cite{BlockSci}, on a 64 GB machine. The final labeled dataset used in numerical experiments consists of $30.331.700$ addresses, associated with $|\mathscr{E}_{known}|=272$ entities representing $5$ entity categories in the following proportions:
\begin{itemize}
    \item \textit{Exchange}: 108 entities, 7.892.587 addresses.
    \item \textit{Service}: 68 entities, 17.606.608 addresses.
    \item \textit{Gambling}: 65 entities, 2.775.810 addresses.
    \item \textit{Mining Pool}: 19 entities, 78.488 addresses.
    \item \textit{DarkNet Marketplace}: 12 entities, 1.978.207 addresses.
\end{itemize}
While the set of labeled address is of significant size, it is important to observe the entity category class imbalance, with the dominant \textit{Service} class representing more than $50\%$ of the dataset, and the smallest \textit{Mining Pool} category representing less than $0.5\%$ of the dataset.

\end{subsection}

\begin{subsection}{Model calibration procedure}

The decision tree model described in Section~\ref{sec:classMeth} is deployed in Python via the LightGBM implementation~\cite{NIPS2017_6907}. The Gaussian Process-based optimization procedure for hyper-parameter optimization is implemented using the Python skopt library\footnote{\url{https://scikit-optimize.github.io/}} with initial parameter values obtained from a coarse random search.

We use a typical $70/30$ training$/$test partition of our dataset.  The learning rate  hyper-parameter of the decision tree model is optimized over the interval $[0.01,0.5]$ after having done a random search over $[0,2]$; the resulting value is  $0.18$. To train LightGBM, we use an early stopping procedure which stops the training if the log loss does not decrease over ten consecutive iterations. The procedure stops after $61$ iterations with a loss of $0.35$. 

The inverse of the $L^{2}$ regularization parameter of the logistic regression model is optimized over the interval $[0.01,3]$ after having done a random search over $[0,5]$;  the value  obtained is $0.55$. 

The Gaussian Processes procedure is used with 50 iterations, which is a reasonable compromise between a small computation time, less than one hour for LightGBM in our hardware setting, and a good exploration of the interval. For fairness we use the same criterion for the logistic regression model. Along the same lines, we only optimize one hyper-parameter for LightGBM, namely the learning rate.
\end{subsection}

\begin{subsection}{Feature importance}

We analyze the performance of the classification model first from the perspective of an unsophisticated attacker incrementally adding features to a generic model based on their ease of access. We then model a sophisticated attacker with extensive modeling knowledge, collecting the full set of features, calibrating model hyper-parameters, and identifying the minimal set of relevant features required for a successful attack.

\begin{subsubsection}{Weak attacker}

Consider an attacker who collects features in the following order, from simplest to most complex:
\begin{itemize}
\item Address features requiring access only to the address set,
\item Entity features requiring access to the address set and address clustering heuristics,
\item Temporal features,
\item Centrality features requiring crawling the Blockchain transaction network for connectivity information,
\item 1,2,3-motif features: requiring crawling the Blockchain transaction network for both connectivity and transaction information (amount, fee, etc.).
\end{itemize}

We  consider both the advanced decision tree model as well as the logistic regression model for classification, representing the range of sophisticated and  simple methods. To model the weak attacker, we do not invoke the hyper-parameter calibration procedure, and use the default parameter settings. F1 and accuracy scores are reported in Table~\ref{table:fe_group_influence}. Given that 1-motifs are simpler to compute in practice for an attacker, we place them before Temporal and Centrality features when reporting the results.
\begin{table}[htb]
\centering
\caption{\textbf{Incremental grouping of features} and associated performance metrics.}
\begin{tabular}{||c c c c c c||} 
 \hline
 Features & |Features| & Alg. & Accuracy & $F_{1}$ & Precision\\ [0.5ex] 
 \hline\hline
 Address & 10 & LR & 0.415 & 0.303 & 0.351\\
 Entity & 18 (+8) & LR & 0.476 & 0.369 & 0.445\\
 1-motif & 62 (+44) & LR & 0.524 & 0.471 & 0.474\\
 Temporal & 78 (+16) & LR & 0.512 & 0.493 & 0.498\\
 Centrality & 120 (+42) & LR & 0.561 & 0.545 & 0.551\\
 2-motif & 201 (+81) & LR & 0.585 & 0.574 & 0.573\\
 3-motif & 315 (+114) & LR & 0.841 & 0.835 & 0.857\\
 Address & 10 & LGBM & 0.5 & 0.487 & 0.492\\
 Entity & 18 (+8) & LGBM & 0.476 & 0.429 & 0.415\\
 1-motif & 62 (+44) & LGBM & 0.622 & 0.597 & 0.613\\
 Temporal & 78 (+16) & LGBM & 0.659 & 0.649 & 0.654\\
 Centrality & 120 (+42) & LGBM & 0.610 & 0.597 & 0.603\\
 2-motif & 201 (+81) & LGBM & 0.683 & 0.654 & 0.667\\
 3-motif & 315 (+114) & LGBM & 0.890 & 0.886 & 0.897\\ [1ex] 
 \hline
\end{tabular}
\label{table:fe_group_influence}
\end{table}
\end{subsubsection}

The results provide two  relevant insights. First it is notable that even a weak attacker using a simple model (logistic regression) with default hyper-parameter settings, and using only the 10 simplest features, is able to identify the entity type with $41\%$ accuracy, i.e. a factor of two improvement as compared to a random guess over the $5$ classes.

Second, it is of significance that while the model choice does not significantly impact the classification performance, using more sophisticated features provides drastic improvement. Indeed, the user  of 3-motif features, encompassing the behavior of the 3-hop graph neighborhood of a given identity, contributes more than $30\%$ relative improvement to the accuracy score (from $0.68$ to $0.89$ accuracy for the decision tree model).

\begin{subsubsection}{Strong attacker}
We now consider a strong attacker collecting the full set of features, and then selecting a small set of highly significant features. We model this process by applying the hyper-parameter calibration procedure described earlier. We then obtain the ranked feature set from the decision tree model. Table~\ref{tablebestparm} provides the top ten features of the decision tree model.
\begin{table}[htb!]
\centering
\caption{\textbf{Ranked features}: by importance according to the decision tree model.}
\begin{tabular}{||c c c||} 
 \hline
 Rank & Type & Name\\ [0.5ex] 
 \hline\hline
 1    &  3-motif   &   unique\_entity\_3\_successor  \\ 
 2    &  Entity   &  prop\_coinbase   \\ 
 3    &  2-motif   &   loop\_2\_std\_nb\_inputs  \\ 
 4    &  2-motif   &   loop\_2\_mean\_nb\_inputs  \\ 
 5    &  3-motif   &   outgoing\_3\_mean\_nb\_outputs  \\ 
 6    &  3-motif   &   outgoing\_3\_std\_nb\_outputs  \\ 
 7    &  3-motif   &   outgoing\_3\_std\_fee\_1\_btc  \\ 
 8    &  3-motif   &   outgoing\_3\_mean\_nb\_inputs  \\ 
 9    &  1-motif   &   incoming\_mean\_fee\_usd  \\ 
 10    & 3-motif    &  loop\_3\_mean\_fee\_2\_btc   \\ [1ex] 
 \hline
\end{tabular}
\label{tablebestparm}
\end{table}

The results  indicate that the 3-hop graph neighborhood features dominate. Note, however that there are 114 such 3-hop graph neighborhood features out of 315 features in total.  

It can be seen from Table~\ref{tablebestparm} that outgoing features are more relevant  that incoming features. This supports a ``causal'' interpretation that features specific to an entity can be well detected from its downstream transaction trace.

We now examine the minimal set of features required to obtain state-of-the-art accuracy. We present in Figure~\ref{fig:ranking_boxplot_small} the F1 and accuracy improvement over the complete feature relevance ranking from the decision tree algorithm for both the logistic regression and the decision tree model.
\begin{figure}[!htb]
    \centering
    \includegraphics[scale=0.5]{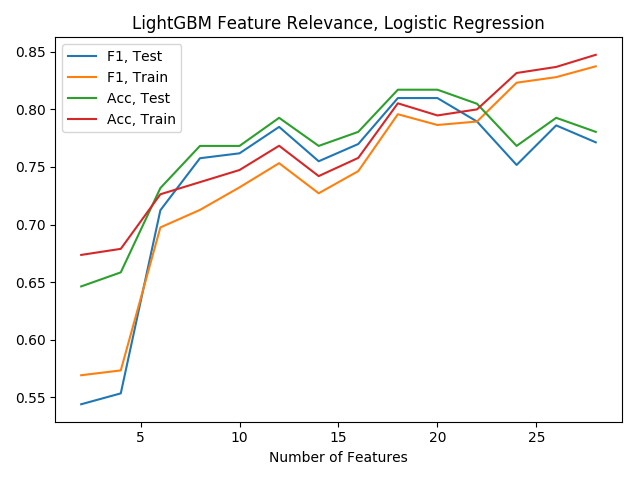}\\
    \includegraphics[scale=0.5]{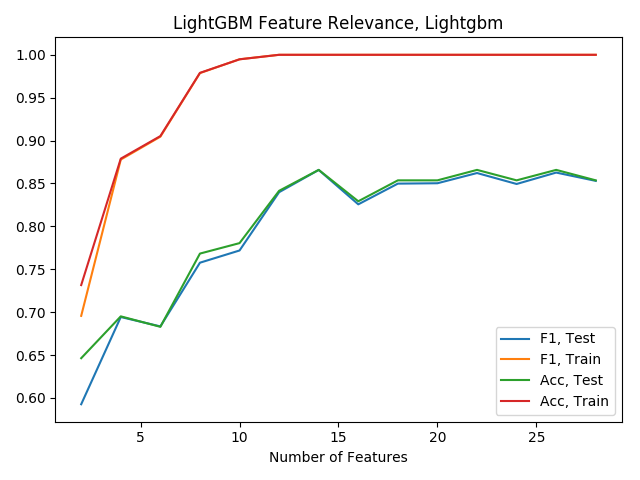}
    \caption{\textbf{Feature Selection} using LightGBM}
    \label{fig:ranking_boxplot_small}
\end{figure}

It is clear from the figure that even with a simple logistic regression model, the $20$ most relevant features are sufficient to obtain high classification accuracy. Using a more sophisticated decision tree model allows reducing the number of features to  $15$, after which improvement plateaus.
\end{subsubsection}
\end{subsection}

\begin{subsection}{Overall model performance}

Finally, we consider the sophisticated attacker using  hyper-parameter optimization. Table~\ref{table:optimized_gp_summary} presents the Accuracy and F1 score for both classification methods.
\begin{table}[htb]
\centering
\caption{\textbf{Optimizing $F_{1}$} with $315$ features, $50$ iterations of the hyper-parameter optimization procedure, global results.}
\begin{tabular}{||c c c c||} 
 \hline
 Algorithm & Accuracy & $F_{1}$ & Precision\\ [0.5ex] 
 \hline\hline
 Logistic Regression  & 0.85 & 0.85 & 0.87\\
 LightGBM & 0.92 & 0.91 & 0.92\\ [1ex] 
 \hline
\end{tabular}
\label{table:optimized_gp_summary}
\end{table}

Table~\ref{table:optimized_gp}, provides the class-specific overall performance results.
\begin{table}[htb]
\centering
\caption{\textbf{Optimizing $F_{1}$} with $315$ features, $50$ iterations of the hyper-parameter optimization procedure, class-level results.}
\begin{tabular}{||c c c c c||} 
 \hline
 Category & Algorithm & Accuracy & $F_{1}$ & Precision\\ [0.5ex] 
 \hline\hline
  Exchange & LR  &  0.91  &  0.91  &  0.91 \\
  Gambling & LR  &  0.9  &  0.82  & 0.75  \\
  Mining & LR  &  0.5  &  0.67  & 1.0  \\
  Service & LR  &  0.85  &  0.87  &  0.89 \\
  Darknet & LR  &   0.75 &  0.75  & 0.75  \\
  Exchange & LGBM  &  0.94  & 0.92   &  0.91 \\
  Gambling & LGBM  &  0.95  &  0.97  &  1.0 \\
  Mining & LGBM  &  0.5  & 0.67   & 1.0  \\
  Service & LGBM  &  0.95  & 0.88   & 0.83  \\
  Darknet & LGBM  &  1.0  &  1.0  &  1.0 \\[1ex] 
 \hline
\end{tabular}
\label{table:optimized_gp}
\end{table}

The results dominate state-of-the-art results from the literature, which may be due to the use of novel advanced graph neighborhood features, as evidenced from Table~\ref{table:fe_group_influence}, along with the hyper-parameter optimization. At the class level, good classification accuracy (above $95\%$) is achieved over the set of exchanges, gambling services and general services as well as the darknet category. Mining pool behavior is less-well captured as evidenced by the  low accuracy, indicating that there is no consistent transaction pattern identified for this class.
\end{subsection}
\end{section}

\begin{section}{Conclusion}\label{sec:conc}
 We formulate the problem of analyzing Bitcoin Blockchain transaction graph anonymity properties as a classification problem over a set of categories of Bitcoin users. Our results indicate that it is feasible for a weak attacker to characterize entities using a set of new graph neighborhood features that we propose, and that is feasible for a strong attacker to do the same with as little as $15$ of the most relevant features. 
 
This suggests a number of interesting avenues for further work. Since it is possible as we have shown to accurately classify transacting entities on the Bitcoin Blockchain, the question arises as to whether it is possible to develop an effective  generative model of the transaction network. If so, it would enable a wealth of studies into the effect of changes in network protocols or regulatory frameworks on the evolution of the Bitcoin economy. 
 \end{section}
\bibliographystyle{plain}
\bibliography{bib_roma}
\end{document}